# Requisite Variety, Autopoiesis, and Self-organization

Carlos Gershenson
Instituto de Investigaciones en Matemáticas Aplicadas y en Sistemas &
Centro de Ciencias de la Complejidad
Universidad Nacional Autónoma de México
cgg@unam.mx
http://turing.iimas.unam.mx/~cgg/

**Abstract**

*Ashby's law of requisite variety states that a controller must have at least as much variety (complexity) as the controlled. Maturana and Varela proposed autopoiesis (self-production) to define living systems. Living systems also require to fulfill the law of requisite variety. A measure of autopoiesis has been proposed as the ratio between the complexity of a system and the complexity of its environment. Self-organization can be used as a concept to guide the design of systems towards higher values of autopoiesis, with the potential of making technology more "living", i.e. adaptive and robust.*

**Complexity**

Cybernetics has studied control in systems independently of their substrate (Wiener, 1948). This has allowed to use the same formalisms to describe different phenomena, such as neuronal and electronic circuits, offering the advantage of allowing the transfer of solutions in one domain to another. For example, understanding adaptive behavior in animals can help us build adaptive machines (Walter 1950; 1951).

The cybernetic tradition of studying systems independently of their substrate has propagated into other fields, one of them being the scientific study of complexity. The term *complexity* derives from the Latin *plexus*, which means interwoven. We can say that a complex system is one in which its elements are difficult to separate, as they depend on each other. This dependence is caused by relevant *interactions* (Gershenson, 2013a). Interactions are relevant when they co-determine the future of elements. Thus, it is not possible to describe the future of an element studied in isolation, its interactions have to be taken into account, and that is why they are relevant. Interactions generate novel information which is not specified in initial nor boundary conditions, limiting the predictability of complex systems (Gershenson, 2013b) because of their inherent computational irreducibility (Wolfram, 2002), *i.e.* you can only know the future once you have been there.

Complex systems are pervasive, as it is more difficult to find isolated phenomena compared to phenomena which interact. It is not that science had not noted the relevance of interactions. We just did not have the proper tools until a few decades ago. It is difficult to describe complex systems in detail with traditional analytic methods, as we have a rather limited number of variables we can fit in a paper or blackboard. The scientific study of complex systems has progressed in parallel with the technological development of electronic computers. This is because computers allow us to study systems with multiple components and their interactions. More recently, data availability has made it possible to contrast different theories of complex systems in all branches of science. In an analogous way to microscopes permitting the study of the microworld and telescopes enabling the exploration of the cosmos, computers are tools which are giving us access to a greater understanding of complexity (Pagels, 1989).

A central question in the scientific study of complex systems is related to their control, given their inherent limited predictability (Gershenson, 2007). It is desirable to predict the future of systems and their environment to be able to act before a perturbation occurs which might damage or destroy the system. In such situations, feedforward control is suitable. Still, depending on the predictability of a system, different control approaches are required (Zubillaga et al., 2014). The less predictable a situation is, the more adaptive a system should be, *i.e.* feedback control would be more appropriate. To be able to decide over different control approaches, measures of predictability and complexity are needed.

In the context of telecommunications, Shannon (1948) defined a measure of uncertainty which he called *information*, which is equivalent Boltzmann's entropy from thermodynamics. Intuitively, a message will carry few information if new data is predictable, *i.e.* it is known beforehand with certainty derived from a skewed probability distribution (few states highly probable). A message will have a high information content for homogeneous probability distributions (most states equally probable), because new data cannot be predicted from data already received. Shannon's information can be used to measure emergence, self-organization, and complexity (Fernández *et al*., 2014).

*Emergence* can be understood as the creation of novel information. For example, interactions between atoms of hydrogen and oxygen generate novel properties in a molecule of water. Interactions between molecules of water generate novel properties such as wetness, temperature, and pressure. The properties at a higher scale are not present in the elements at the lower scale, so we can say that the properties emerge from the interactions between elements. In other words, emergence occurs when interactions generate novel information. Shannon's information can be used precisely to measure this novel information, and thus emergence, and normalized to the interval between zero and one. Maximum emergence will occur with maximum information, *i.e.* minimum predictability, while minimum emergence will occur with minimum information, *i.e.* maximum predictability.

*Self-organization* occurs when interactions between elements of a system produce a global pattern or behavior, as in the stripes of a zebra of a flock of birds. There is no central or external control, but local interactions between elements lead to global regularities. Maximum organization is achieved with maximum regularities. Thus, self-organization can be seen as the inverse of entropy, and thus information and emergence (Pask and von Foerster, 1960). Self-organization will be high when emergence is low and vice versa. A maximum self-organization occurs with minimum information, *i.e.* maximum predictability, while minimum self-organization occurs with a maximum information, *i.e.* minimum predictability.

*Complexity* requires both emergence and self-organization. Following López-Ruiz *et al*., (1995), we can define complexity as the multiplication between emergence and self-organization. Thus, complexity will be minimal when emergence (chaos) or self-organization (order) are extreme, and complexity will be high when there is a balance between emergence and self-organization. Control of systems with a high complexity will benefit from having both prediction and adaptation. For a formal derivation and further explanation of these measures of emergence, self-organization, and complexity, please refer to Fernández *et al*. (2014).

**Requisite variety**

Ashby's *law of requisite variety* states that an active controller must contain as much variety as the phenomenon it attempts to control (1956). This is because if the controlled has, say, ten states, the controller must be able to have enough variety to respond appropriately for each of those ten states. For example, if an industrial robot should be able to install four different types of screws, it requires enough variety to discriminate at least four different situations and act accordingly to each of them. If we now require the robot to manage two new types of screw, it will require a greater variety to consider six different situations than the one required in the four screw scenario.

Stafford Beer applied the law of requisite variety to organizations, highlighting that systems need to have enough variety to be *viable*. Still, in some cases the environment of systems may have "insufficient" variety, and controllers should attempt to increase the variety of the environment rather than matching it (Espejo and Reyes, 2011).

Bar-Yam (2004) noticed that variety can be seen as a synonym of complexity, which had not been defined in its current usage when Ashby proposed his law. This law can be generalized to a *law of requisite complexity*: an efficient active controller will require at least the same complexity as the complexity of the controlled. In other words, a controller for a complex system requires to be at least as complex as the system it attempts to control. In practice, this requires a balance between predictability and adaptability of the controller (Gershenson, 2013b) to face both the emergence and self-organization of the controlled.

Living systems can be described as control systems. Thus, the law of requisite variety also is found in life, as organisms have to match the complexity of their environment at different scales.

**Autopoiesis**

Maturana and Varela (1980) coined the concept of *autopoiesis* to define living systems. Autopoiesis means self-producing, so it is related to autonomy (Ruiz-Mirazo and Moreno, 2004). Even when the original concept focussed on the emergence of biological systems from chemical components, the use of autopoiesis has been generalized to other domains, such as sociology (Luhman, 1986).

A measure of autopoiesis has been proposed recently (Fernández, *et al*., 2014), taking inspiration from the law of requisite variety and the concept of "life ratio" (Gershenson, 2012a). Autopoiesis can be defined as the ratio between the complexity of a system and the complexity of its environment. Note that the environment of a system is not the rest of the universe, but only the part which interacts with it, i.e. affects the system (von Uexküll, 1985). If autopoiesis is less than one, it implies that the environment has a higher complexity than the system, thus dominates its dynamics. If autopoiesis is greater than one, it means that the system has a greater complexity than its environment, *i.e.* it fulfills the law of requisite complexity. An autopoiesis greater than one also implies a higher degree of autonomy of the system. This generalized view of autopoiesis considers systems as self-producing not in terms of their physical components, but in terms of its organization, which can be measured in terms of information and complexity. In other words, we can describe autopoietic systems as those producing more of their own complexity than the one produced by their environment. Thus, a greater (informational) autopoiesis implies more autonomy.

Notice that autonomy is gradual: having a certain autopoiesis does not mean that a system is independent of its environment. This suggests that life is also a gradual property: the transition from non-living to living is smooth (Gershenson, 2012a). Moreover, the concept of life can be generalized to domains other than biology as well (Langton, 1997; Aguilar, *et al*., 2014). In this view, we can build systems with the features of the living (Bedau, *et al*., 2009; 2013). But how to do it?

**Guided self-organization**

We can use self-organization to guide the dynamics of complex systems (Prokopenko, 2009; Ay *et al*., 2012; Polani *et al*., 2013; Prokopenko, 2014; Prokopenko and Gershenson, 2014). It was already mentioned that self-organization consists of an internal increase in order of a system. Since most systems have certain emergence (entropy) for free (thermodynamics), self-organization can be used guide systems towards a higher complexity (Gershenson, 2012b), and thus autopoiesis and variety.

*Guided self-organization* can be understood as the steering of the self-organizing dynamics of a system toward a desired configuration. We can assume that environmental variety (emergence) is given for free, because of the second law of thermodynamics. Therefore, we can focus on applying self-organization to match the particular (required) variety to control the environment. Few emergence will require few self-organization, as there will be few variety. High emergence will demand a high self-organization to match the high variety. One approach for designing and controlling self-organizing systems consists on minimizing friction (negative interactions) and maximizing synergy (positive interactions) (Gershenson, 2007) through the implementation of *mediators*. These mediators must match the required complexity of the environment to be able to cope with the variety of different possible interactions and states which might occur. Moreover, this has to be done at multiple scales (Gershenson, 2011), as complexity is dependent on scale (Bar-Yam, 2004).

An example where these ideas have been applied is in the coordination of traffic lights (Zubillaga *et al*., 2014). Comparing a self-organizing method (Gershenson, 2005) with a traditional non-adaptive method, we have shown that the self-organizing method is close to a theoretical optimal performance for all densities (Gershenson and Rosenblueth, 2012). This is achieved because the controller (traffic lights) manages to adjust its complexity to the complexity of its environment (vehicles), leading to an autopoiesis greater than one for almost all densities. The densities where autopoiesis is less than one is precisely where the performance is farther from the optimum.

A similar approach can be generalized to other domains (Gershenson, 2007), such as cognitive systems (Haken and Portugali, 2015) or urbanism (Gershenson, 2013c). Most urban systems are complex because the interactions between their elements generate novel information (emergence) at different temporal and spatial scales. This limits their predictability, while urban problems change in time, *i.e.* they are non-stationary. Self-organizing urban systems can adjust their complexity to match the variety of the urban problems as they change in time, thus maintaining an autopoiesis greater than one, as it occurs with traffic lights. In this sense, we can speak about cities becoming more living (Gershenson, 2013c).

**Discussion**

The approach presented so far can also be used to describe and understand the evolution of complexity (Gershenson and Lenaerts, 2008). If we assume random variations in information, some information will be able to propagate better. This implies that information able to cope with the complexity of its environment will have a higher probability of persistence, leading to the natural selection of more complex information. Since this information will be part of the environment of other systems, this will push other systems towards an increase in their complexity as well. In this sense, complexity and life as understood here is to be expected in evolution with very few assumptions considered (Gershenson, 2012a).

Life as a systemic property, more general than the original concept of autopoiesis (Froese and Stewart, 2010). It might be useful to use a more abstract conception of life because it allows us to study the properties of living systems beyond biology. Since it is desirable to have these properties to face the complexity around us, from a pragmatic perspective we can say that such a description can be useful.

## Conclusions

Systems will be viable if their complexity (variety) is higher than the complexity of their environment. This is also true for living systems. Thus, we can define a measure of autopoiesis as the ratio between the complexity of a system and the complexity of its environment. To achieve higher complexities, self-organization can be used to guide the properties and dynamics of systems towards a balance to match the emergence imposed by the environment.

Such a general formulation can be rather abstract, but is useful to direct efforts to design and control complex systems. Its benefits have been already shown for urban systems. There is a potential to be explored in other domains, where artificial systems can be designed to be more like living systems.

## Acknowledgements

I am grateful to Mikhail Burtsev, Raúl Espejo, Nelson Fernández, Roberto Murcio, and Jesús Siqueiros for valuable comments and suggestions.